\documentstyle[12pt]{article}
\begin{document}
\title
{\bf The Chirality operators for Heisenberg Spin Systems}
\author{
V. Subrahmanyam\cr
International Center for Theoretical Physics, P.O. Box 586\cr
34100 Trieste, Italy\cr}
\maketitle
\begin{abstract}
The ground state of closed Heisenberg spin chains with an odd number of sites
has a chiral degeneracy, in addition to a two-fold Kramers degeneracy.
A non-zero chirality implies that the spins are not coplanar, and is a measure
of handedness. The chirality operator, which can be treated as a spin-1/2
operator, is explicitly constructed in terms of the spin operators, and is
given as commutator of Permutation operators.
\end{abstract}
\hoffset =-1truecm
\parindent=0.5in
\baselineskip =0.8truecm
\newpage
\vskip 1in
In this report, we investigate the chirality operator for Heisenberg spin
chains. Usually the chirality operator is given in terms of fermion operators
defined\cite{kn:Wen89} through
${\vec s}_i\equiv c_{i\alpha}^{\dagger}{\vec \sigma_{\alpha\beta}}
c_{i\beta}$, and the fermion hopping operator $T_{ij}=\sum_{\alpha}c_{i\alpha}
^{\dagger}c_{j\alpha}$. The chirality operator is given as the imaginary
part of difference of products of $T$ operators along the two closed loops on
a closed chain. Unscrambling the fermion operator to express the chirality
operator in terms of the original spin operator can be tedious in general,
though the answer for the case of three spins turns out to be very simple,
a box
product of the three spins ${\vec s_1}.{\vec s_2}\times {\vec s}_3$. Below
we give an explicit construction of the chirality operator in terms of the
spin operators directly. We treat the chirality as
a spin-1/2 operator ${\vec \chi}$, whose z-component is the usual
chirality operator discussed above.
The construction
starts with defining the chiral-spin raising operator $\chi^+$ and its
hermitian conjugate, and $\chi^z$ is given by the commutator
$2 \chi^z=[\chi^+,\chi^-]$, from the standard spin-1/2 algebra. The
the chiral raising and lowering operators are related to
permutation operators (permutation operators naturally arise
as the chirality can be changed by a permutation of the labels), which
can be expressed explicitly in terms of the original spin operators.

The Heisenberg Hamiltonian is given by
$$
H=\sum_{i}{\vec s}_i.{\vec s}_{i+1}
$$
where the sum is over $N$ sites of a closed chain. We consider the case of an
$N=3$ as it is the simplest case, and generalize for any odd $N$ later.
The ground state for an odd-numbered chain belongs to the sector with the
total spin $S=1/2, S^Z=\pm1/2$, implying a two-fold Kramers degeneracy.
However, each of the sectors with $S^z=\pm1/2$ is further two-fold degenerate.
This extra degeneracy is due to chirality. The two chiral states in $S^z=1/2$
sector can be written in basis as
$$
\phi_{\pm}= \sum_{l} {\rm e}^{\pm ik(l-1)} |l>
$$
where $k=2\pi/3$ and $|l>=s_l^-|\uparrow\uparrow\uparrow>$. The
two ground states in $S^z=-1/2$ sector are obtained by operating on the
states above with the total spin lowering operator.
The two chiral
states have spin currents going in two different directions. For $N\ne3$, the
spin current corresponds to the motion of the center of mass of the down
spins.
We refer to these two states as having chirality by defining $\chi^z|\phi_\pm>
=\pm {1\over2}|\phi_\pm>$. Similarly the chiral-spin raising operator, along
with its hermitian conjugate,  is
defined through $\chi^+|+>=0, \chi^+|->=|+>$.
We give the explicit form of the chirality operator ${\vec\chi}$
in terms of the spin operators below. This can be done with the help of
the permutation operators.

Let the permutation operator $P_1$ denote the permutation $23\rightarrow
32$. Similarly the operators $P_2, P_3$,
permute 3 and 1, and 1 and 2
respectively.
These permutation operators acting
on one of the chiral states give the other chiral state modulo a phase.
This implies that we should be able to
construct the $\chi^+$ and $\chi^-$ in terms of the $P$ operators.
The $P$ operators are constructed easily from the original spin operators,
for instance $P_1=2({\vec s}_2.{\vec s}_3+1/4)$ and so on.
The action of $P_1$ on the two chiral states can be seen easily,
$P_1|+>=|->$ and $P_1|->=|+>$.
This implies $P_1=\chi^++\chi^-=2\chi^x.$ Similarly we can construct the
other two operators, $P_2=\kappa^2\chi^+\kappa\chi^-$ and $P_3=\kappa\chi^+
+\kappa^2\chi^-$, where $\kappa=\exp(ik)$.
It should be noted that the $P$ operators are linearly dependent $P_1+P_2+
P_3=0$, and they satisfy interesting commutator relations
\begin{equation}
\chi^z = {1\over 4i\sin{k}} [P_i,P_{i+1}]
\label{eq:perm}
\end{equation}
for all $i$, and the anticommutator is -1. In fact this can be used as the
definition of the chirality operator ${\vec \chi}$, in conjunction with
$\chi^x=P_1/2$, and $i\chi^y= [\chi^z ,\chi^x]$. Writing back the permutation
operators in terms of the spin operators, we recover the box product form
for $\chi^z$. The advantage of writing ${\vec \chi}$ in terms of the
permutation operators is that Eq.\ref{eq:perm} holds for any general $N$.
The only difference is now $k=2\pi/N$, and the permutation operators should
be suitably constructed, which we illustrate below.

Before we come to the generalization of Eq.\ref{eq:perm}, we present a simple
understanding of the two-fold chiral degeneracy of Heisenberg chains using
the Jordan-Wigner fermion
representation of the spin operators\cite{kn:Sch66}. Let $s_i^+=c_i^{\dagger}
\exp{i\pi
\sum_{j<i}n_j}$, and $s_i^z=n_i-1/2$, where $c_i$ is a fermion annihilation
operator at site $i$ and $n_i$ the fermion number operator. Under this
transformation the xy-part of the Heisenberg transforms into a fermion hopping
term and the z-part becomes a nearest-neighbour interaction for the fermions.
Let us focus on the xy-terms alone, as it turns out that the z-part of the
Hamiltonian does not change the chiral degeneracy. The xy-part of the
Heisenberg Hamiltonian reduces to $H_{xy}=J/2\sum_{i=1}^{N-1}(c^{\dagger}_i
c_{i+1} +h.c.)+(-1)^{N_F+1}(c_1^{\dagger}c_N+h.c.)$. The periodic boundary
condition of the closed chain gives rise to phase factor for the hopping
amplitude between the sites 1 and $N$, which depends on the number of
fermions $N_F=S^z+{N\over2}$. The eigenstates of the above Hamiltonian
are plain wave states, and the eigenvalues come in doublets except the lowest
(or the highest depending on whether $N_F$ is odd or even) eigenvalue. If we
fill $N_F$ states from below for $S^z=1/2$ sector, for $S^z=-1/2$ the states
get filled from above, and vice versa. In both cases the highest occupied
doublets wil
have only one fermion, which gives a degeneracy of two. Also there is a fermion
current, and as the two eigenfunctions are related by a phase, the current
changes direction between the two states. If we include the z-part of
the Heisenberg Hamiltonian, the single-particle picture we have now does
not hold. However, the single-particle current will translates into the
motion of the center of mass, which means in the spin language that the center
of mass of down (or up depending on the spin sector) spins moves around the
chain in two directions, with a phase of $\exp{\pm ikx_{cm}}$, where
$x_{cm}$ is the coordinate of the center of mass. This can be seen explicitly
for $N=5$, from the Bethe anstaz states\cite{kn:Sub93}, though for general
$N$ it is quite involved.

With the above insight that the two chiral states differ by the phase
of the motion of the center of mass of the down spins, we can now construct
the permutation operators appearing in Eq.\ref{eq:perm}.
For a given site
$i$ we have an operator $P_i$ which permutes the spin labels such that
$i+j\rightarrow N+i-j$. This is equivalent to doing a reflection on a regular
polygon with $N$ sites, around a straightline bisecting the angle at site $i$.
For instance in the case of $N=5$, the operator $P_1$ permutes the spin labels
$12345\rightarrow 15432$, and $P_2$ does the permutation $12345\rightarrow
32154$ and so on. The $P$ operators can be readily constructed from the
spin operators, as each of them involve $(N-1)/2$ pair-wise permutations, as
\begin{equation}
P_i=2^{N-1\over2} \prod_{j=1}^{N-1\over2} (s_{i+j}.s_{N+i-j}+{1\over4}).
\label{eq:spin}
\end{equation}
We need to relate these permutation operators to the chiral-spin raising and
lowering operators, which will give us the desired construction of the
chiral operator $\vec \chi$ in terms of the spin operators.
Let us label the spins on a closed chain such that
$P_1|\pm>=|\mp>$, implying $P_1=\chi^++\chi^-$, as we have seen in the case
of three spins. The action of the other $P$ operator can be seen from the
action of the cyclic permutation operator $R$, which permutes the spin
labels such that $i\rightarrow i+1$. The $R$ operator shifts all the spin
labels by one unit, which gives a shift of one unit for the center of
mass of the down spins, $i.e.$ $R|\pm>=\exp(\mp ik)|\pm>$. Let
$|\tilde +>=\kappa^*|+>, |\tilde ->=\kappa|->$, where $\kappa=\exp{ik}$.
The action of $P_2$ on these
new states $|\tilde \pm>$ is similar to the action of $P_1$ on $|\pm>$,
$i.e. P_2|\tilde \pm>=|\tilde \mp>$. This gives us
$P_2=\kappa^*\chi^++\kappa\chi^-$. Now using the operator $R$ again, we can
construct $P_3$, and the procedure can be repeated to get the other operators.
In a concise notation we have $P_i$ given as
$$
P_i=A_i^* \chi^+ + A_i \chi^-,
$$
where $A_i=\kappa^i$.
The commutator relation between $P_i, P_{i+1}$ remain the same the same as in
the
case of three spins as given in Eq.\ref{eq:perm}. More generally we
have
\begin{equation}
[P_l,P_{l+m}]=4i \chi^z\sin{mk},
\label{eq:gen}
\end{equation}
and the anticommutator is 2$\cos{mk}$.
This completes the construction
of the chirality operators, Eq.\ref{eq:spin} and Eq.\ref{eq:gen}, in terms of
the spin operators.
It is
interesting to note that for each of the permutation operator $[P_i,H]=0$,
this is because the Hamiltonian is invariant under an anticyclic permutation.
This implies the chirality operators, as they involve only
the permutation operators explicitly, commute with the Hamiltonian.

Now we would like to comment on the closed chain with an even number of sites.
For even $N$ the ground state of Heisenberg antiferromagnet belongs to $S=0$
sectors, and is nondegenerate. It is also easy to see from the Jordon-Wigner
transformation we discussed above. For $N=2n$, the number of fermions is
$N_F=n$. For $n$ even the
eigenvalues spectrum for the $xy-$part consists of $n$ doublets, and the ground
states is nondegenerate with $n/2$ doublets filled in. For odd $n$ the
eigenvalue spectrum consists of nondegenerate highest and lowest energy states
and the rest of the states in doublets, and the ground state is again
nondegenerate with $(n-1)/2$ doublets and the lowest nondegenerate
state filled in.
This implies $\chi^z=0$. However this situation is not
analogous to the $S=3/2$ excited states of a triangle with $\chi^z=0$. The
difference is that for the case of three spins (or for any odd $N$) the
chirality
operator exists, and still there could be states with $\chi^z=0$, while for
for even $N$ the chirality operator does not exist, $i.e.$ $\chi^z\equiv 0$.
For example consider $N=4$, which is the simplest nontrivial case. It is easy
to see that $P_1=P_3$, as in both cases only one pair of spin labels 2 and 4
have to be permuted. We have $[P_1,P_2]=[P_2,P_3]$, implying $\chi^z\equiv 0$.
This argument can be readily generalized for any even $N$. Noting that
$P_i=P_{{N\over2} +i}$, from the commutator relations of Eq.\ref{eq:gen}
for $\lbrace l=1,m={N\over2}-1
\rbrace$ and
$\lbrace l={N\over2}, m=1\rbrace$, we have $\chi^z=-\chi^z$, which implies
the chirality does not exist.

In summary, we have constructed the chirality operators for the two-fold
chiral ground states of Hiesenberg antiferromagnetic chains with odd number of
sites explicitly
in terms of the spin operators. It would
be interesting to see if chirality operators can be constructed
independent of the interaction between the spin degrees of freedom.

\parindent=0.5in
\noindent{\bf Acknowledgement}

It is a pleasure to thank A. M. Sengupta for discussions.

\end{document}